\documentclass[]{article}
\usepackage{epsfig}
\usepackage{emulateapj}
\usepackage{apjfonts}
\usepackage{stuff}

\begin{document}

\slugcomment{Accepted for publication in the Astrophysical Journal (Letters)}
\revised{May 11, 1999}
\received{March 23, 1999}
\def\MR{M\'esz\'aros and Rees}
\title{Gamma-ray Burst Spectral Features: 
Interpretation as X-ray Emission from a Photoionized Plasma}
\author{C. J. Hailey \altaffilmark{1}, F. A. Harrison\altaffilmark{2} and K. Mori\altaffilmark{1}}

\altaffiltext{1}{Columbia Astrophysics Laboratory,
Columbia University, New York, NY 10027}
\altaffiltext{2}{Space Radiation Laboratory and Division of Physics, Mathematics
and Astronomy, California Institute of Technology, Pasadena, CA 91125}

\begin{abstract}
Numerous reports have been made of features, either in emission or
absorption, in the 10 - 1000 keV spectra of some gamma-ray bursts.
Originally interpreted in the context of Galactic neutron star models
as cyclotron line emission and $e^+ - e^-$ annihilation features,
the recent demonstration that the majority of GRBs lie at
cosmological distances make these explanations unlikely.  In
this letter, we adopt a relativistic fireball model for cosmological
GRBs in which dense, metal rich blobs or filaments of plasma are
entrained in the relativistic outflow.  In the context of this model,
we investigate the conditions under which broadband features, similar to
those detected, can be observed.  We find a limited region of
parameter space capable of reproducing the observed GRB
spectra. Finally, we discuss possible constraints further high-energy
spectral observations could place on fireball model parameters.
\end{abstract}


\section{Introduction}

Numerous detections have been reported of features, either in emission
or absorption, in the spectra of some gamma-ray bursts (GRB).
Absorption features below 100 keV were reported by
Konus~(\cite{mga+81}), {\em HEAO A-1} (\cite{hueter84}), and {\em
Ginga} (\cite{mfh+88}), and in addition Konus and other instruments
have detected broad emission-like features at high energy (400 -- 500
keV) in a few cases~(e.g. \cite{mgi+79,tc80}).  The BATSE Spectroscopy
Detectors (SD) on {\em CGRO} have recently reported 13 statistically
significant line candidates, although some uncertainty in the
contribution of systematics to the analysis make these detections
uncertain (\cite{bbp+99}).  For the most statistically significant
BATSE candidate, GRB 930916, the feature appears as a broad bump
between 41 and 51~keV(\cite{bbp+99}).  Even if the BATSE features are
not confirmed, the SD data cannot yet rule out the existence of
features similar to those seen by {\em Ginga} in some fraction of GRBs
(\cite{brf+96}), and confirmation awaits more sensitive spectroscopic
gamma-ray instruments.

Originally, these line features were interpreted in the context of 
Galactic neutron star models for the GRB progenitors. The low-energy
absorption features were explained as cyclotron resonance lines in a
$\sim 10^{12}$ Gauss magnetic field (\cite{hl90,fce+88}), and the
high-energy lines were postulated to be from $e^+ - e^-$ annihilation
radiation gravitationally redshifted near the surface of a solar mass
neutron star (e.g. \cite{liang86}).  Recently, however, detection of
redshifted absorption lines in the optical counterparts associated
with two bursts (\cite{mdk+97},\cite{kdo+99}),
and emission lines from the
galaxies associated with three others (\cite{kdr+98},\cite{dkg+98},\cite{d+99})
have confirmed cosmological distances for five
GRBs. Although the BATSE data still allow a fraction of GRBs to be in
a Galactic distribution (\cite{lw95}), the majority of long GRBs must be
cosmological.  It is possible that the observed gamma-ray spectral features are
associated with a subpopulation of Galactic GRB progenitors.  However,
given the recent redshift and host galaxy observations this seems
unlikely.  It is therefore interesting to look for an explanation for
these features in the context of cosmological GRB models.

Recently, M\'esz\'aros and Rees (1998)\nocite{mr98} have discussed the
possibility that the relativistic outflows associated with
cosmological GRBs may entrain small blobs or filaments of dense,
highly-ionized metal rich material that could give rise to broad
features due to Fe K-edges in the GRB spectrum.  For typical blob
Lorenz factors of $\Gamma_b \sim 25 - 100$, Fe K-edges would give rise
to isolated broad features in the 250 -- 1000~keV band, similar
perhaps to the high-energy lines observed by Konus.  It would be
difficult, however, to produce the line observed by {\em Ginga}, which
have multiple features below 100~keV.  In this letter, we accept the
line detections as real, and investigate the possibility that the
low-energy lines seen by {\em Ginga} could be produced by excitation
and absorption in the predominantly Ne-like Fe-L complex and/or in
outflows containing low-Z elements.  As we show in \S\ref{results},
features similar to the {\em Ginga} lines can be produced both by O,
Ne, Si -rich outflows, and by a combination of Fe and light elements.
Finally, in the context of this model, we discuss the utility of broad
band X- and gamma-ray spectroscopy for constraining fireball model
parameters and the composition of the ejecta.

\section{Model and Calculations}\label{model}

In the model described by \MR\ (1998) (hereafter MR98),
metal-enriched, high-density regions become entrained in the fireball,
and are confined by the high ambient and ram pressure of the
relativistic outflow. These small blobs or filaments, although a negligible
fraction of the total outflow mass, can have a significant covering
factor.  Blobs with gas temperature comparable to the comoving photon
temperature ($\sim$1 keV for typical fireball model parameters) would
form photoionized plasmas with prominent line emission, similar to
those found in many X-ray emitting sources.  The plasma density would
be high by most astrophysical standards, reaching $n_b \sim
10^{18}$cm$^{-3}$ (assuming typical fireball parameters) for the case
where the blob internal pressure balances the total external (magnetic
and particle) pressure. These blobs would be accelerated by radiation
or magneto-hydrodynamic pressure and would achieve a saturation bulk
Lorentz factor well before reaching the emission region where internal
shocks convert a significant fraction of the bulk kinetic energy into
radiation ($r_{sh} \sim 10^{13}$cm).

As described in MR98, several factors would broaden any spectral
features from the dense photoionized plasma.  If all blobs
have the same bulk Lorentz factor, $\Gamma_b$, emission line features
will be broadened due to contributions from regions with velocities
that are at different angles to our line of sight, which will have
different Doppler blue-shifts.  In addition, the blobs may have a
range of Lorentz factors, since those with low enough column would be
accelerated to the velocity of the surrounding flow, whereas those
with larger columns would reach slower terminal velocities.  The range
of Lorentz factors then depends on the range of blob sizes, which in
turn depends on the details of the entrainment process and the extent
to which instabilities break up the bigger blobs.  

We have adopted this basic scenario described in MR98, and
investigated the conditions under which low-energy features similar to
those seen by {\em Ginga} can be formed.  Our goal was to
qualitatively reproduce multiple spectral features with a similar
fraction of the total luminosity.  It is important to note that the
actual shape of the {\em Ginga} features is very sensitive to proper
subtraction of the underlying continuum, and therefore on proper
understanding of the instrument response.  Given the imperfect
knowledge of the {\em Ginga} response, varying assumptions about the
continuum can cause the features to appear qualitatively different
(\cite{fce+88}).   We assume that the blobs are in
pressure equilibrium with the surrounding medium (otherwise they would
not be stable), and in addition, treat the optically thin regime only.
Although the model itself does not impose any limit on the optical
depth, the optically thin assumption represents the simplest case,
where self-shielding and other time-dependent effects can be ignored.

We used the photoionization code, XSTAR (\cite{kk95}), to directly
calculate the reprocessed spectrum and temperature of the photoionized
material. As input to the code, we must specify the ionization
parameter, $\Xi = L/(n_b r^2)$ (here $L$ is the luminosity, $n_b$ the
blob particle density, and $r$ the blob size).  In the comoving frame,
$\Xi = L/n_br^2\Gamma_b^2$ (MR98).  Additional inputs are the ionizing
spectrum and relative elemental abundances.  For the ionization
parameter, we investigated the range $\Xi = 100 - 1000$ consistent
with the fireball model parameters of MR98.  Similarly we considered
the relevant range $\Gamma_b = 25 - 100$.  We assume a power-law
ionizing spectrum with energy index $\alpha$, varied between 0.1 and
0.5, consistent with the average continuum spectrum early in the
burst.  Few constraints can be placed on the relative elemental
abundances, since for neutron star mergers or hypernova scenarios,
little is known about the composition of the surface.  We have
therefore investigated a range for the abundant elements O, Ne, Si,
and Fe.

The XSTAR output, after ionization equilibrium is established,
consists of the blob temperature (in the comoving frame), the
abundance of each ion, prominent lines and edges with location and
magnitude, and the ratio of bolometric recombination line to continuum
luminosity.  Given the assumption of pressure equilibrium, the choice
of $\Xi$ fixes the blob temperature in the co-moving frame; $\Xi = 500
T_7$, where $T = 10^7 T_7$~K (MR98, assuming an isotropic blob
distribution).  We discard as inconsistent any solutions that do not
satisfy the required relationship between $\Xi$ and $T$.  We also
consider a range of reprocessing rates (ratio of total recombination
luminosity to total ionizing luminosity), which we adjust in order
to find solutions with a ratio of deposited luminosity in the
broad features to total luminosity of a few percent, consistent
with the {\em Ginga} observations.  We keep only solutions
that have total optical depth in the lines, $\tau \leq 1$,
consistent with the optically thin assumption.  

With the output from XSTAR, and the assumed reprocessing rate, we
calculate an observed spectrum by blueshifting (by $\Gamma_b$), and
broadening with the instrumental width and the (dominant) relativistic
effect resulting from the variation of velocity projected along the
line of sight.  To determine the magnitude of the latter, we assume
the Lorentz factor and the luminosity to be independent of time, and
we integrate over the spherical emission surface, assuming the
emitting material to be uniformly distributed.  This results in a
spectral smearing of $\sim$50\%, similar to the 30 -- 50\% suggested
by MR98.  In addition, we investigated the effects of time-varying
luminosity by parametrizing a decreasing luminosity resulting from
shell expansion, and calculating the line broadening over the entire
shell.  For this time-dependent calculation, we included proper
integration over the equal arrival time ellipse, as described by
Panaitescu and M\'esz\'aros (1998)\nocite{pm98}.  The time-dependence
does result in additional broadening, however it is not sufficient to
qualitatively change our conclusions, and for simplicity we therefore
employ the time-independent calculation in the results presented here.
We do not include any additional broadening due to possible range of
blob bulk Lorentz factors.

The density of the blobs is very high by normal astrophysical
standards, and the validity of the XSTAR code under these conditions
is therefore of concern.  XSTAR ignores three-body recombination,
assuming that the ionization equilibrium is determined by
photoionization, radiation, and dielectronic recombination.  Explicit
calculation shows that due to the high temperatures, radiative
recombination will dominate over three-body recombination, and this
will not result in significant inaccuracy.  A further concern is that
the code does not properly treat collisional redistribution among
excited levels. We estimate these errors to be at the $\sim$25 --
100\% level, and not of concern for reproducing gross spectral
features.

Finally, we note that our results are not strongly dependent on the
geometry of the entrained material.  \MR\ (1998) point out that the
blobs may have a filamentary structure resulting from the magnetic
fields, however given our assumption of optically-thin emission,
this will not significantly alter the observed spectrum.  

\section{Results}\label{results}

By investigating the range of parameter space described above, we
found that we could reproduce spectral features in the 10 -- 100 keV
band resembling the BATSE and {\em Ginga} measurements.  Figures~1 and
2 show results from two of the best cases (i.e. those most closely
resembling the broad features seen by these instruments).  For each,
we show the spectrum of reprocessed photons in the comoving frame, the
spectrum after relativistic broadening, and the observed spectrum
after convolving with the {\em Ginga} instrument resolution.  We have
not included any additional broadening due to possible spread in the
blob Lorentz factors.  The first case (Figure~1) has a relative
abundance of metals of O:Ne:Si = 1:0.5:1, and the second (Figure~2)
has O:Ne:Si:Fe = 1:0.25:0.15:0.05.  The values for the ionization
parameter and energy spectral index are indicated in the captions.  To
reproduce features with the intensity of the {\em Ginga} or BATSE
detections requires a reprocessing rate on order unity ($\tau \sim
1$), marginally consistent with the assumption of optically thin
emission.

\begin{figure*}
\centerline{\epsfig{file=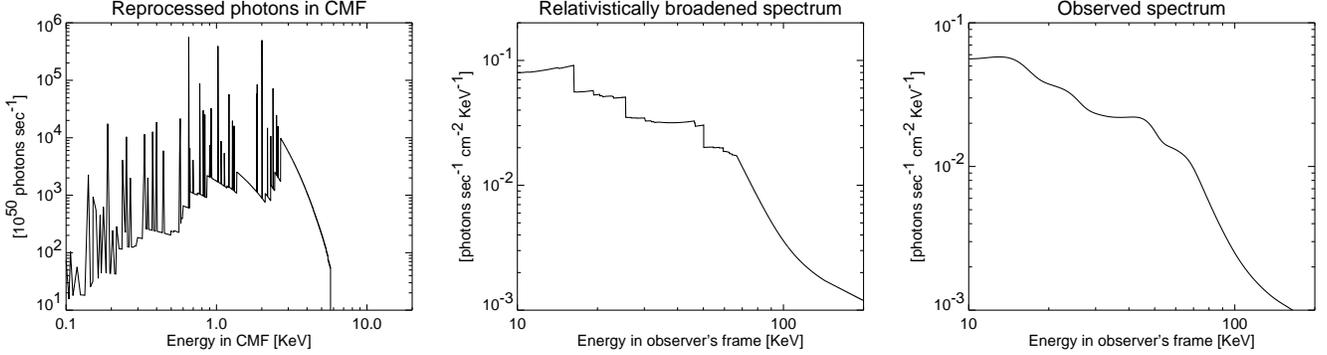,angle=90,width=18cm}}
\caption{The panels, from left to right, show the
spectrum of reprocessed photons in the comoving frame, the spectrum
including reprocessed and continuum flux after relativistic
broadening, and after convolution with a response function typical of
{\em Ginga}.  This case has a relative elemental abundance of O:Ne:Si
= 1:0.5:1, log($\Xi$) = 2.1, a power law energy index $\alpha = 0.1$
for the ionizing flux, a plasma temperature in the comoving frame 
$T_{cmf} = 7.1 \times 10^6$~K, and a reprocessing rate of unity.  The resulting
spectrum has $\sim$2\% of the luminosity in the two dips between 10 - 100 keV,
similar to the {\em Ginga} detections. For spectral normalization we
have assumed $z = 1, \Omega = 0.2, H_o = 65$, and
total GRB luminosity $L = 10^{51}$ erg/s.}
\label{ons}
\end{figure*}

\begin{figure*}
\centerline{\epsfig{file=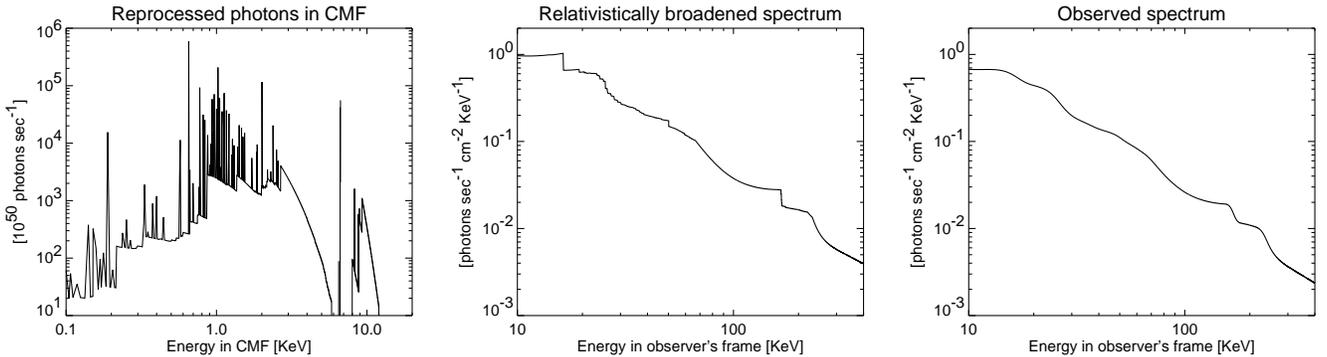,angle=90,width=18cm}}
\caption{The same as in Figure~1, except that the
relative metal abundance is O:Ne:Si:Fe = 1:0.25:0.15:0.05,
log($\Xi$)=2.3, $\alpha = 0.3$, and $T_{cmf} = 7.3 \times 10^6$~K.
The two features appearing above 100 keV are due to Fe K emission
($\sim$150 keV) and the Fe K edge ($\sim$220 keV).  These features
could be interpreted as those observed by {\em ISEE}.}
\label{onsf}
\end{figure*}

From Figures 1 and 2, it is clear that multiple features resembling
absorption dips can be produced below 100 keV from a combination of
low-Z elements, and in the case where Iron is present, from the
L-shell complex.  These dips, interpreted as absorption features in
the {\em Ginga} spectra, are a result of smearing of the complex
ionization structure (line emission and edges), and are not due to
absorption.  If the ejecta contain Iron (Figure 2), then additional
emission-like features are seen in the 100 -- 200 keV band due to the
K-shell.  Note that only one instrument has reported high-energy lines
simultaneously with low-energy features. This is due primarily to
instrumental limitations: reasonably large collecting area is required
for high-energy detection, while good energy resolution and clean
instrumental response is required below 100 keV, and these have not
been combined in a single experiment.  We have therefore adjusted the
blob Lorentz factor to the value required to match the {\em Ginga}
observations ($\Gamma_b = 25$) for a GRB with redshift $z = 1$.  This
value of the Lorentz factor is smaller than the $\Gamma_b \sim 100$
typically invoked to ensure the emission region is not opaque due to
photon-photon pair production.  The latter value is, however,
estimated by assuming the gamma-ray spectrum extends to 100~MeV. There
is no evidence for such high-energy emission in the {\em Ginga}
events, where the low-energy line features were observed, and the
value $\Gamma_b = 25$ is consistent with all the observations.
Note that the 300 - 500 keV emission-like features
seen in some GRB could be associated with Fe K emission edges for
values of $\Gamma_b \sim 50 - 100$.  Including only broadening due to
the variation of velocity projected along the line of sight and
typical instrumental resolution produces features consistent with the
observations, while any additional broadening due to variation in blob
Lorentz factors would smear out the lines entirely.  We emphasize that
the results presented here are the best cases, achieved by searching a
fairly wide range of parameter space.  For many conditions still
consistent with reasonable fireball model parameters, no observable
features are produced.

\section{Conclusion}

We have investigated the possibility that the 10 -- 100 keV features
reported in the spectra of some GRB arise from smearing of the
reprocessed radiation from a metal-rich, dense photoionized plasma
entrained in blobs or filaments in the relativistic outflow.  By
searching a relatively wide region of parameter space consistent with
generic fireball models, we can reproduce the observed features for a
limited set of values and elemental composition. The dips in the
spectrum observed after relativistic and instrumental broadening are
due to the complex ionization structure, and are a combination of
emission lines and edges. In addition, the sum of the recombination
spectrum plus continuum produces a break in the observed continuum at
$\sim 100$ keV, similar to that seen by BATSE.  With no additional
broadening due to a range of $\Gamma_b$, relativistic effects are not
sufficient to smear out the features in the recombination spectrum
entirely.

If such dense, entrained blobs do exist, the broadband spectral
features can be used to constrain the range of fireball model
parameters, as well as the composition of the entrained material.  In
particular, the presence of Iron in the blobs would produce features
due to K shell emission in the 100 - 1000 keV band that could be used
to measure the Lorentz factor of the ejecta.  We emphasize that
detailed comparison with existing observations is not possible due to
the uncertainties in the instrument response, and the poor signal to
noise of the detections.  Small variations in the response function,
as well as assumed continuum, can severely alter the characteristics
of the observed features (i.e. whether they are interpreted as
emission or absorption dips).  The primary characteristics of an
experiment capable of confirming and measuring these spectral features
is broad energy response (few keV -- 1 MeV), large area, and clean,
well-determined response function, and moderate energy resolution.

\acknowledgements{The authors wish to thank Masao Sako for assistance
with XSTAR, and William Goldstein for useful discussions on the
atomic physics of photoionized plasmas.}





\end{document}